\documentstyle[12pt]{article}

\begin{document}

\newcounter{eq}[section]
\newcommand{\set}{\stepcounter{eq}
\renewcommand{\theequation}{\mbox{\arabic{section}.\arabic{eq}}}}

\hsize=6.15in
\vsize=8.2in
\hoffset=-0.42in
\voffset=-0.3435in

\normalbaselineskip=24pt\normalbaselines

\begin{center}
{\large \bf How does connectivity between cortical areas depend on \\
brain size? Implications for efficient computation}
\end{center}

\vspace{0.15cm}

\begin{center}
{Jan Karbowski}
\end{center}

\vspace{0.05cm}

\begin{center}
{\it Sloan-Swartz Center for Theoretical Neurobiology,
Division of Biology 216-76, \\
California Institute of Technology,
Pasadena, CA 91125, USA \/}
\end{center}


\vspace{0.1cm}

\begin{abstract}
A formula for an average connectivity between cortical areas in mammals
is derived. Based on comparative neuroanatomical data, it is found,
surprisingly, that this connectivity is either only weakly dependent or
independent of brain size. It is discussed how this formula can be
used to estimate the average length of axons in white matter.
Other allometric relations, such as cortical patches and area sizes vs.
brain size, are also provided. Finally, some functional implications, 
with an emphasis on efficient cortical computation, are discussed as well.
\end{abstract}





\vspace{1.5cm}

Email: jkarb@cns.caltech.edu, jkarb@its.caltech.edu

\vspace{1cm}

Journal Ref: Journal of Computational Neuroscience 15, 347-356 (2003)

\newpage

\section{Introduction}

The salient feature of a macroscopic cortical organization is the
presence of different functional units such as columns and areas. These 
units form an ordered network of connections mediated by axonal
bundles in white matter. On a microscopic scale, however, neurons are 
connected in sparse, stochastic local circuits (Braitenberg, 1978a;
Douglas and Martin, 1991). The transition between
the two levels of organization takes place on a column-size scale.

It has been argued that both microscopic and macroscopic connectivity,
i.e., the fraction of connected sites, should decrease with brain size 
(Stevens, 1989; Ringo, 1991; Ringo et al, 1994), because this would decrease 
the total axonal length; a feature 
desirable especially in bigger brains. This expectation follows from 
the so-called minimal axon length principle, which assumes that the brain 
tries to save its biochemical resources (Mitchison, 1992; Cajal, 1995; 
Cherniak, 1995; Murre and Sturdy, 1995; Chklovskii and Stevens, 1999).

The purpose of this article is to investigate whether this is a justified
presumption by studying how both microscopic and macroscopic connectivity
depend on brain size. It is found that, although the connectivity
between neurons decays with brain size (a known fact), the
connectivity between cortical areas is either weakly decreasing
or invariant with the brain volume. We argue that the formula for the
connectivity between areas can be used to estimate an average axon
length in white matter, thus providing a useful practical tool, especially
in the face of lacking data. We also suggest that brains, in order to
perform efficient computation, have to sacrifice some of their computational
power, and this is due to a trade-off between managing brain size and its
limited biochemical resources, and maintaining functional operativeness. 
We speculate that this trade-off may lead to observed scaling laws between 
some cortical parameters.

In all considerations below, it is assumed that the brain volume $V_{b}$
and the gray matter volume $V_{g}$ scale with an exponent close to 1, 
which is well
justified experimentally (Jerison, 1973; Frahm et al, 1982; 
Prothero and Sundsten, 1984; Hofman, 1985; Hofman, 1989; 
Barton and Harvey, 2000). Therefore,
we use terms ``scale with gray matter volume'' and ``scale with brain
volume/size'' interchangeably and treat them as equivalent.

For completeness, first let us determine how neuronal
connectivity depends on brain size.
An average connectivity $p$ (or probability of connection) between neurons 
can be defined as $p= M/N$, where $M$ is the average number of
synapses per neuron, and $N$ is the total number of neurons in the
cortex. Since the volume density of synapses in 
gray matter is brain size independent (Sch{\"u}z and Demianenko, 1995; 
Braitenberg and Sch{\"u}z, 1998),
i.e. $NM/V_{g}= const$, then we
obtain that $M\sim V_{g}/N$, and as a consequence $p\sim V_{g}/N^{2}$.
The total number of neurons $N$ is proportional to the total cortical
surface area $W$ (Rockel et al., 1980). The latter scales with the brain
volume as: $W\sim V_{g}^{0.9}$ for large convoluted brains (Hofman, 1985). 
This leads to the following scaling 
between the average connectivity $p$ and brain size for convoluted brains

\set
\begin{equation}
p\sim V_{g}^{-0.8}. 
\end{equation}\\
Thus $p$ decreases quickly with brain size. As an example, the human brain 
volume and the rat brain volume differ by a factor of $614.6$ (Hofman, 1985).
From this, it follows that the average connectivity $p$ is about $170$ times
smaller in human than in rat. 

In the next Sections we study scaling of the area connectedness
with the brain volume.

\vspace{0.2cm}

\noindent {\bf List of symbols used in this article}

\noindent $a$ is a dimensionless parameter characterizing cortical geometry
and a pattern of axonal organization in white matter,

\noindent $c$ is the ratio of the volumes of gray matter and the whole brain,

\noindent $d$ is the average diameter of axons in white matter, 

\noindent $E$ is the basal metabolic energy rate used by gray matter at rest,

\noindent $f$ is the fraction of active excitatory synapses in gray matter 
at resting conditions,

\noindent $K$ is the number of areas in the cortex,

\noindent $L_{0}$ is the average length of axons in white matter,

\noindent $M$ is the average number of synapses per neuron in the cortex,

\noindent $N$ is the total number of neurons in the cortex,

\noindent $<n>$ is the average degree of separation between cortical areas, 

\noindent $p$ is the average probability of connection between neurons 
in the cortex,

\noindent $q$ is the probability of sending at least one macroscopic axonal
bundle to white matter by a module, 

\noindent $Q$ is the average  probability of connection between two areas,

\noindent $P_{n}$ is the probability that two areas are connected in at
least $n$ steps,

\noindent $S_{n}$ is the probability of connection of two areas via at least
one of the paths that uses $n$ intermediate areas (steps), 

\noindent $W$ is the total cortical surface area,

\noindent $W_{0}$ is the surface area of one cortical area,

\noindent $V_{g}$ is the gray matter volume,

\noindent $V_{w}$ is the white matter volume,

\noindent $V_{b}$ is the brain volume,

\noindent $V_{body}$ is the volume of the whole body,

\noindent $\alpha$ is the scaling exponent of the cortical area number with the gray
matter volume,

\noindent $\beta$ is the scaling exponent of the axonal length in white matter
with the gray matter volume,

\noindent $\gamma$ is the scaling exponent of the white matter vs gray matter
volume,

\noindent $\delta$ is the scaling exponent of the cortical area connectedness
with the gray matter volume,

\noindent $\kappa$ is the probability of connection between a given module 
in one area to some other area, 

\noindent $\tilde{\kappa}$ is a fraction of connected cortex by a module,

\noindent $\xi$ is the linear size of a module in the cortex,

\noindent $\tau$ is the average conduction delay between cortical areas.

\vspace{0.2cm}

\section{Connectivity between cortical areas}

In this Section a probability of a connection between cortical areas
is derived. We assume that the cerebral cortex of the total surface
area $W$ is divided into $K$ areas of an identical surface area $W_{0}$.
Each of the areas, in turn, is composed of modules of linear size $\xi$.
We define a module as a local group of neurons with similar functional
properties that is capable of sending at least one macroscopic (containing
at least several axons) coherent bundle of long-range 
(cortico-cortical) axons to
a particular place in the cortex (Pandya and Yeterian, 1985; Zeki and 
Shipp, 1988; Felleman and Van Essen, 1991). One can think about 
such defined modules as being cortical columns or barrels that have been
found in the visual and somato-sensory cortices of different mammals.

Consider two cortical areas A and B. We assume that the area A is
connected to the area B if at least one of the modules in A is connected
to B. Note that this definition of connectivity ignores the issue of
strength of the connection, and focuses only on the existence of
a link between areas. Also, we neglect the explicit influence of the
distance between areas on their connectivity, since no precise quantitative
anatomical data exists. In this sense, we consider only ${\it average \/}$
area connectedness. However, we include implicitly the spatial effect 
by considering the average length of axons connecting different areas. 
If we assume that there are limited biochemical resources
in the cortex, then in particular, the length of axons that modules
can send to white matter should be kept as small as possible. 
This restriction puts
a spatial constraint on the number of areas a module can reach, and this
in turn, influences the probability of connection between areas.
An average probability $\kappa$ that a given module in A
connects with any module in B (probability of connection between a module in A
and the whole area B) is given by

\set
\begin{equation}
\kappa= q(W_{0}/W)(a L_{0}^{2}/W), 
\end{equation}\\
where $q$ is the
probability that the module in A sends at least one macroscopic axonal 
bundle to white matter. This probability should be close to 1, since
e.g. for the mouse cortex pyramidal cells, many of which
project to white matter, constitute $85 \%$ of the total number of cortical
cells (Braitenberg and Sch{\"u}z, 1998) and a module consists of about
$10^{3}$ neurons (Braitenberg, 1978b). Thus, it is very likely that 
a sufficient number of axons
is sent to form at least one coherent bundle, and this process should not 
depend on brain size.
The ratio $W_{0}/W$ in eq. (2.1) is the probability that this bundle 
terminates in the area B. In general, this probability can be different
for a different pair of cortical areas (c.f. Young et al, 1995) and does
not have to be equal to $W_{0}/W = 1/K$. However, since we are interested 
only in average probability, the assumption of uniformity is reasonable
and sufficient. 
The dimensionless ratio $a L_{0}^{2}/W$ is the probability that the 
area B can be
physically reached given finite average axonal length $L_{0}$ in 
white matter (fraction of areas that can be connected given finite fiber
range). The factor $a$ is some dimensionless constant characterizing 
a particular
cortical geometry (either convoluted or smooth) and a pattern of 
axonal organization in white matter (e.g., how many, on average, bundles
a module sends to white matter). This factor can be weakly 
species-dependent, however, it should not depend in any systematic manner 
on brain size. We do not make any specific
assumptions about the cortical or axonal geometry, and therefore our results 
and conclusions drawn are quite general. The latter do not depend on a
particular value of the parameter $a$.

The probability $Q$ that the area A connects with the area B (area connectedness) 
is exactly complementary to the probability that none of the modules in A 
connects with the area B, i.e. $(1-\kappa)^{W_{0}/\xi^{2}}$,
where $W_{0}/\xi^{2}$ is the number of modules in every area. 
Thus, our $Q$ is given by

\set
\begin{equation}
Q= 1 - (1-\kappa)^{W_{0}/\xi^{2}}.
\end{equation}\\
We can simplify this formula by using an identity 
$1-\kappa = \exp[\ln(1-\kappa)]$, and by expanding the logarithm in it
for small $\kappa$ according to $\ln(1-\kappa) \approx -\kappa$.
This produces: $(1-\kappa)^{W_{0}/\xi^{2}} 
\approx \exp(-\kappa W_{0}/\xi^{2})$. The next step is to substitute the
right-hand side of eq. (2.1) for $\kappa$, and to use the fact that
$W_{0}= W/K$. After these operations, we finally obtain the following 
expression for the average connectivity $Q$:

\set
\begin{equation}
Q\approx 1 - \exp\left(- \frac{a qL_{0}^{2}}{\xi^{2}K^{2}}\right).
\end{equation}\\
From this formula, which is one of the main ones in this paper, it follows 
that the average area connectedness
for a given cortex depends on its four basic characteristics: cortical
geometry, the average length of axons in white matter, module size, 
and the number of cortical areas.

It is also possible to find an expected number of modules in one area
that connect with another area. Assuming that modules are statistically
independent, i.e. the probability of sending axonal bundles for
a given module does not depend on other modules, the distribution of
the number of modules in A reaching B
is represented  by a binomial distribution. Thus the average number of
modules in A connecting with B is given by the product  of the probability
that a module in A connects with area B ($\kappa$) and the 
number of modules
in A ($W_{0}/\xi^{2}$), i.e. $aqL_{0}^{2}/\xi^{2}K^{2}$. 
The latter expression 
is exactly the same as the argument in the exponent in eq. (2.3), and 
this fact provides a useful interpretation of that equation.

To find how $Q$ scales with brain size, one therefore first has 
to determine
how the expected number of connected modules between two areas
depends on the brain volume. This is equivalent to finding how
the above basic characteristics scale with brain size.
This will be done in the next section. 
Here, we estimate the average connectivity $Q$ for the mouse cortex,
and also, the average length of long-range axons for few other species.
All the numbers provided are for one isolated hemisphere only.

Mouse is the only animal for which the distribution of axonal length in 
white matter of one hemisphere has been measured directly 
(Greilich, 1984; Braitenberg and Sch{\"u}z, 1998, Fig. 62). 
From this, one can estimate that 
$L_{0}\approx 3$ mm. Module size $\xi$ can be assumed to be of the
order of an average barrel size, i.e. $\xi^{2}\approx 0.1$ mm$^{2}$. 
The unknown factor $a$ in eq. (2.3) can be estimated
from data of Sch{\"u}z and Liewald (Sch{\"u}z and Liewald, 2001). They found,
using injections of anterograde tracer BDA, that neurons under a surface area
of about $0.1$ mm$^{2}$, i.e. of approximately a barrel size,
in the mouse cortex project onto $12 \%$ of the total cortical surface area
of one hemisphere.
The corresponding number for the macaque cortex is about $1-2 \%$
(Sch{\"u}z, private communication). To find $a$ we use eq. (2.1)
for the probability of connection between a module in A and the 
area B. In the present case, however, the surface area of B is actually
the whole cortex (of one hemisphere), i.e. $W_{0}= W$, and we obtain the 
following equation:

\set
\begin{equation}
\tilde{\kappa} = aqL_{0}^{2}/W,
\end{equation}\\
where $\tilde{\kappa}= 0.12$ is the fraction of mouse
cortex which is connected by a module.
The total cortical surface area is a ratio
of the gray matter volume, which in one hemisphere is $V_{g}= 56$ mm$^{3}$, 
and cortical thickness,
which is about $0.85$ mm (Braitenberg and Sch{\"u}z, 1998). Combining all the 
numbers, we obtain $a\approx 0.87$ (for $q\approx 1$), which can be used 
to determine the area connectedness.
The number of cortical areas for mouse depends on the methodology 
used (see below). If we use the Caviness criteria (Caviness, 1975) then
$K= 24$. However, if we use methodology of Kaas et al, then $K= 6-8$
(Krubitzer and Huffman, 2000).
In the former case, this leads to $Q= 0.13$, while in the latter this yields
$Q= 0.71-0.89$.

For a few other mammals, i.e. rat, cat, and macaque monkey, it has been
possible to determine, based on experimental data, the average connectivity 
$Q$. Unfortunately, we are unable to compare those numbers with predictions
of eq. (2.3) due to the lack of data for $L_{0}$ in these species.
However, we can perform a reversed computation and evaluate the axonal
length for each of these animals directly form eq. (2.3) [i.e. solve eq. (2.3) for 
$L_{0}$], that is

\set
\begin{equation}
L_{0}\approx \xi K \sqrt{ \frac{1}{aq}\ln\left(\frac{1}{1-Q}\right)}.
\end{equation}\\
This formula, as well as formula (2.4) can have practical importance 
for estimation of average
length of fibers in white matter, since it is technically easier to
determine the connectivity $Q$, or the fraction $\tilde{\kappa}$ of connected 
cortex, than to measure directly $L_{0}$. However, the area and 
connectivity counts, must be done within a single methodology (see below).

For all the above animals the factor $a$ and the sizes of modules are 
roughly equal (this will be justified in the next section) to the 
corresponding numbers for mouse.
Additionally, for rat we can find the number of connections between
areas based
on a connectivity graph in Kolb (Kolb, 1990; p. 26, Fig. 2.1). 
The total number of connections between areas  
(we assumed that all of them are reciprocal, which is a
reasonable assumption, although there is no information about this in
the article) is found to be 168, which gives $6.46$ connections per area.
Since the number of areas is
$K= 26$ (Kolb, 1990), we estimate that the connectivity 
for the rat cortex in one hemisphere
is $Q= 6.46/26 = 0.25$. Using eq. (2.5), this yields $L_{0}= 4.7$ mm. 
For the cat cortex with $Q=0.27$ and $K=65$ in one hemisphere
(Scannell and Young, 1993; Scannell et al, 1995; Young et al, 
1995), we obtain $L_{0}= 12.4$ mm. Finally, for macaque 
with $Q=0.15$ and $K=73$ (Young, 1992; Young, 1993; Young et al, 1995), 
we obtain axonal length $L_{0}= 10.0$ mm. 
The latter value, which is somewhat smaller than expected
may result from, for instance, underestimation of the connectivity
$Q$ (Young et al, 1995, p. 130) or the parameter $a$. However, we can
perform a second, independent, evaluation of $L_{0}$ using eq. (2.4). 
Taking the macaque cortical surface area of one hemisphere as $W= 1.25\cdot 10^{4}$ 
mm$^{2}$ (Hofman, 1985; Stephan et al, 1981),
we find $L_{0}= 12.0-17.0$ mm (for $\tilde{\kappa}= 0.01-0.02$; 
Sch{\"u}z, private communication), which seems to be more reasonable.

\section{Scaling of basic cortical characteristics and area connectedness
with brain size}

In this Section we study how the number of cortical areas and their size,
module size, the long-range axon length, and the area connectedness depend on 
the brain volume.

First, let us consider the scaling of the number $K$ of cortical areas
with brain size. Although, there is no consensus on how to define
areas, and different physiological criteria lead to different counts,
it is possible to obtain such a scaling law if one focuses on a single 
methodology.
This has been done by Changizi (Changizi, 2001), who used the methodology
of Kaas and Krubitzer (Kaas, 1987; Krubitzer, 1995; Krubitzer et al, 1997).
He found for 11 species with brain volumes spanning 2 orders of magnitude
that $K\sim V_{g}^{\alpha}$ with $\alpha\approx 0.4$. It is rather unlikely
that another methodology would change this exponent drastically.
It is also interesting to note that a similar scaling law has been proposed
theoretically for hypothetical cortical compartments forming a completely 
connected network (Braitenberg 1978b; Braitenberg, 2001). 

We can combine the above scaling law for the number of cortical areas $K$ 
with the previously presented (in Introduction) scaling law for the total 
cortical surface area $W$ to find a useful allometric relation between
the average cortical area size $W_{0}$ and the brain volume. Since 
$W\sim V_{g}^{0.9}$ (Hofman, 1985), and $K\sim V_{g}^{0.4}$ (Changizi, 2001),
and because $W_{0}= W/K$, we obtain $W_{0}\sim V_{g}^{0.5}$. Thus, the
size of cortical areas increases moderately with brain size.

The dependence of a module size on the brain volume can be determined by
noticing that the module size should correlate strongly with the size of 
cortical patches. The dependence of the size of patches
in V1 of different species on the gray matter volume $V_{g}$ is shown
in Fig. 1. This figure 
reveals that there is no systematic dependence of $\xi$ on $V_{g}$, and this
suggests that $\xi$ is invariant with respect to brain size.

As was mentioned above, there are no data on the average length of long-range
axons in mammals (with mouse being an exception), and therefore we have to 
make a reasonable assumption about its scaling with brain size. 
If we assume that the average axon length $L_{0}$ in white matter
is proportional to its diameter, then $L_{0}\sim V_{w}^{1/3}$, 
where $V_{w}$ is the white matter volume. 
The white matter volume is related to the gray matter volume via an allometric
scaling law: $V_{w}\sim V_{g}^{\gamma}$,
with $\gamma= 1.22-1.33$ (Frahm et al, 1982; Hofman, 1989; Allman, 1999;
Zhang and Sejnowski, 2000; Barton and Harvey, 2000).
It seems that the difference between $1.22$ and $1.33$ in the exponent
$\gamma$ results from the fact that different researchers included
different species in their analysis. For example, the white matter volume 
of insectivores scales against the rest of the brain volume with a lower
exponent than a corresponding exponent for primates (Barton and Harvey, 2000).
Combining the above, we obtain for the average axon length in white matter
$L_{0}\sim V_{w}^{1/3}\sim V_{g}^{\gamma/3} \equiv  V_{g}^{\beta}$, where 
$\beta= \gamma/3$. Thus the exponent $\beta$ is 
in the range $0.41-0.44$. We can verify this result independently.
In Fig. 2 we plot our values of $L_{0}$,
estimated in Sec. 2 based on eqs. (2.4) and (2.5), against the gray matter
volume. The figure yields the value of the exponent $\beta$ about 0.26, 
which is clearly different from the value $\sim 0.4$ above. The source of
the discrepancy between the two numbers may be twofold. First, the
assumption about the linear relationship between axonal length in white 
matter and the white matter diameter may not be quite correct. Second,
our Fig. 2 uses only 4 mammals and that may be a reason for a poor
statistics. It is possible that the true value of the exponent $\beta$ 
lies in the range $0.26-0.44$.

Now we are in a position to determine how the connectivity between 
cortical areas scales with brain size. Combining the scaling laws
for $L_{0}$, $\xi$, and $K$, we find that the expected number of 
connected modules
between two areas $aqL_{0}^{2}/(\xi K)^{2}$ scales as

\set
\begin{equation}
\frac{aqL_{0}^{2}}{\xi^{2} K^{2}}= A\ V_{g}^{2(\beta-\alpha)},
\end{equation}\\
from which we obtain the scaling relation
for the area connectedness

\set
\begin{equation}
Q\approx 1 - \exp\left(-A V_{g}^{-\delta}\right),
\end{equation}\\
where $\delta= 2(\alpha-\beta)$ and
$A$ is some positive brain size independent constant. 
It is apparent
that for the scaling of $Q$ with brain size, it matters only the
relative scaling of the average axon length and the number of areas.
In the case when $\beta\approx 0.4$ the value of the exponent 
$\delta\approx 0$, suggesting almost perfect independence of $Q$ of 
brain size. However, in the second case when $\beta= 0.26$, there should
be a slow decay of $Q$ with the brain volume, since then $\delta= 0.28$. 
Below, we estimate the magnitude of this decay.

The constant $A$ appearing in eqs. (3.1) and (3.2) can be determined by using 
our previous estimates of $a= 0.87$ and $\xi^{2}= 0.1$ mm$^{2}$, data in Fig. 2
on $L_{0}$ scaling, and data on $K$ scaling (Changizi, 2001). From 
Fig. 2 we obtain that $L_{0}= 1.10\ V_{g}^{0.26}$. From Fig. 3 in
Changizi (Changizi, 2001) we find the area number $K= 0.42\ V_{b}^{0.4}$, 
where $V_{b}$ is the brain volume of one hemisphere expressed in mm$^{3}$
(note that Changizi uses the total brain volume of two hemispheres). 
However, the cortical area count 
used by Changizi (Kaas-Krubitzer methodology) underestimates the area
number by a factor of 2.6-4.0 (calculation for monkey, cat, and mouse) 
in comparison to more traditional approaches
(Caviness, 1975; Kolb, 1990; Young et al, 1995). In order to be consistent
with the previous sections, where we used those traditional approaches
in our estimates of the axonal length, we rescale the Changizi scaling law 
by a factor of 3.3, which is the average of the above numbers. 
Thus, a modified 
expression for the $K$ scaling takes the form: 
$\tilde{K}\approx 3.3 K\approx 1.45\ V_{b}^{0.4}$, where $\tilde{K}$
is the modified cortical area number. We also need the relationship between
the brain volume $V_{b}$ and its gray matter volume $V_{g}$ that takes the 
form:
$V_{g}/V_{b}= c$, where $c$ is approximately constant within a single mammalian
group. For example, $c$ is about $0.5$ for primates and $0.1$ for insectivores 
(Stephan et al, 1981; Barton and Harvey, 2000). Inserting all the above into 
eq. (3.1) we arrive at the value of $A\approx 5.0 c^{2\alpha}$.
Thus the scaling relation for the cortical area connectedness in the case
when $\beta= 0.26$ takes the from:

\set
\begin{equation}
Q\approx 1 - \exp\left(-5.0 c^{2\alpha} V_{g}^{-\delta}\right),
\end{equation}\\
where the exponents $\alpha= 0.4$, $\delta= 0.28$, and $V_{g}$ is the 
gray matter 
volume of one hemisphere expressed in mm$^{3}$. As an example, for 
the mouse cortex 
with $V_{g}= 56$ mm$^{3}$ (Braitenberg and Sch{\"u}z, 1998) and since 
for rodents
$c\approx 0.15$ (Stephan et al, 1981), we obtain $Q= 0.30$. 
For the human cortex with $V_{g}= 3.4\cdot 10^{5}$ mm$^{3}$ 
and $c\approx 0.5$ 
(Stephan et al, 1981), we obtain $Q= 0.08$. Although the brains of these
two species differ by almost 4 orders of magnitude, their area connectedness
differs only by a factor of 3.8.

Another characteristic of the cerebral cortex related to its connectivity 
is the
so-called degree of separation between cortical areas. This quantity is defined
as an average minimal number of steps (i.e. number of intermediate areas) which
are necessary to connect two given areas. The detailed calculation of this 
quantity
is presented in the Appendix. Here, we only note that the average degree 
of separation
between areas is slightly less or around $2$ regardless of brain size.

\section{Discussion}

One of the main objectives brains try to accomplish, is to process various 
forms of information. This goal can be achieved by possessing a certain level 
of computational power, which can be controlled by many factors.
Some of these factors may involve molecular processes, some cellular,
and some may involve interactions with the environment. To study all
of them, it is an extremely complicated task. However, if one
focuses only on a coarse-grained macroscopic outcome of these processes,
then a few functional/architectonic principles can be identified which
brains should meet to efficiently process information. 
We suggest that there are three basic principles, which one can view 
as constraints imposed by some hypothetical ``perfect'' design. We will argue
below that deviations from this design in real brains are the consequence 
of compromises that brains have to face.

The first principle is that the number of cortical areas should increase
as quickly as possible with brain size (Kaas, 1995; Kaas, 2000). 
The result of Changizi (Changizi, 2001) on the number of areas vs brain size
is consistent with this assertion.
That trend would allow bigger brains to perform many sophisticated tasks in 
appropriately specialized locations, i.e. locally. This principle implicitly 
assumes that the number of areas is a major contribution to animal's 
capabilities (Kaas, 1995).   
The second principle is that brains try to maintain a constant connectivity
between cortical areas regardless of brain size; this is what the
results of Sec. 3 may suggest. The third principle, which is closely related
to the second, is that intra- and inter-hemispheric temporal delay should
not increase with the brain volume. The last two requirements prevent isolation
of cortical areas and additionally enhance the efficiency of information
transfer between them, thus providing the link between local and global
information processing (Sporns et al, 2000). Although, it has been argued before
that delays do increase with brain size (Ringo et al, 1994), we treat the third
principle as a first order approximation of what one may naively expect to be
a perfect design. The reason for this is that it is intuitively natural to 
assume that delays would somehow interfere with efficient cross-talk between 
areas.

It is interesting to realize that by fulfilling the functional/architectonic
requirements brains would have to deal with an excessive increase in size of 
white matter 
in relation to gray matter. This is undesirable, because this would lead
indirectly to longer cortico-cortical axons, and that would contradict
the principle of minimal axon length 
(Mitchison, 1992; Cajal, 1995; Cherniak, 1995; Murre and Sturdy, 1995;
Chklovskii and Stevens, 1999). 
Below we analyze how this excessive scaling of white matter vs. gray matter 
arises.

If we assume that the cortical white matter is composed 
primarily of cortico-cortical fibers then the white matter volume $V_{w}$ 
is given by proportionality

\set
\begin{equation}
V_{w}\sim NL_{0}d^{2},
\end{equation}\\
where $N$ is the number of neurons in the cortex and $d$ is the axonal
diameter in white matter.
There has been some debate about whether actually the axonal diameter in 
white matter changes with brain size. The early comparison of fibers 
for mouse and macaque (Jerison, 1991; Sch{\"u}z and Preissl, 1996) 
indicated that 
majority of axons in both species have roughly the same diameter, with the 
exception of a small fraction of thick fibers which were present only in 
macaque.
This led to the impression that the fiber diameter can depend only very
weakly on brain size, however these data were not sufficient to determine
the scaling exponent. Recently, the scaling exponent has been determined
for the group of 6 mammals with volumes spanning almost 3 orders of
magnitude using a single methodology (Olivares et al, 2001).
It was found that the average diameter of axons in corpus callosum
scales against the gray matter volume as $d\sim V_{g}^{0.066}$, which 
indeed is quite weak. 

We want to express the white matter volume in terms of 3 functional parameters:
the number of areas $K$, the average connectivity between areas $Q$, and the 
conduction delay 
$\tau$ between areas. For this we have to find how $L_{0}$ and $d$ depend
on these parameters. $L_{0}$ is related to $K$ and $Q$ via eq. (2.3). If we
expand the exponent in that equation and retain only the leading order term, then
$Q\sim (L_{0}/K)^{2}$. Thus, approximately $L_{0}\sim K Q^{1/2}$. 
The conduction delay $\tau$ between areas scales as $\tau\sim L_{0}/v$, 
where $v$ is the velocity of signal transmission in the white matter axons. 
Since, most of the 
axons in the white matter are myelinated, the velocity $v$ is proportional to the 
first power of axonal diameter $d$, that is, $v\sim d$ (Hursh, 1939; Rushton, 1951).
Thus, the conduction 
delay $\tau\sim L_{0}/d$. From which, it follows that 
$d\sim L_{0}/\tau \sim K Q^{1/2}/\tau$. Finally, from the Introduction, we
have that the number of neurons $N\sim V_{g}^{0.9}$. 
Combining all the above, we find that the ratio of volumes of white and 
gray matters depends on the functional parameters in the following
way:

\set
\begin{equation}
V_{w}/V_{g} \sim V_{g}^{-0.1} \frac{K^{3} Q^{3/2}}{\tau^{2}}.
\end{equation}\\
This equation shows that, if the above three hypothetical functional 
principles 
were satisfied then the white matter volume would have to grow excessively
with the gray matter volume due to fast growth of $K^{3}$ with brain size. 
This would lead to an undesirable situation when, above certain brain scale, 
fibers require more space and possibly relatively more biochemical resources
than the units processing information.
A more optimal situation would be to require slower increase of $K$ with brain
size and simultaneously to allow $\tau$ to increase, and $Q$ to decrease 
slowly with brain size, because
this would reduce a scaling exponent of the right-hand side of eq. (4.2)
with $V_{g}$. That would probably slightly decrease information-processing 
capabilities of the brain but on the other hand, it would also decrease its 
size and biochemical costs.
Such strategy seems to be taken by brains, since approximately
$\tau\sim L_{0}/d \sim V_{g}^{0.20}$ and $Q\sim V_{g}^{-0.28}$ (if we take
the exponent $\beta= 0.26$). 
Thus, in order to manage their biochemical resources and size brains
probably have to sacrifice some of their computational power by choosing 
compromised scaling exponents. Note that we reached a similar conclusion 
regarding conduction delays as Ringo et al (Ringo et al, 1994), but using 
a different perspective.

Another constraint imposed on brains, indirectly related to the above 
principles, is associated with metabolic processes in neurons during 
information transfer (Laughlin et al, 1998). One can expect that for an
efficient coding the metabolic energy consumption should be low 
(Levy and Baxter, 1996; Karbowski, 2001; Karbowski, 2002). Below, we
present a mathematical argument which is consistent with this hypothesis.

Imaging techniques have been pointing out that glutamatergic excitatory 
synapses are the major users of metabolic energy in the brain 
(Wong-Riley, 1989; Sibson et al, 1998; Shulman and Rothman, 1998). 
Recent estimate of the
distribution of energy expenditure among different processes confirms
that conjecture, especially for bigger brains (Attwell and Laughlin, 2001).
Thus, the metabolic energy $E$ used by gray matter should be roughly 
proportional
to a number of active excitatory synapses in its volume. Since the proportion
of excitatory and inhibitory synapses in the cortex seems to be brain size
independent (Sch{\"u}z and Demianenko, 1995), one can conclude that the
number of excitatory synapses should be proportional to the total number of
synapses $NM$ in the cortex regardless of brain size. This implies that

\set
\begin{equation}
E\sim f N M, 
\end{equation}\\
where $f$ is the fraction of active excitatory synapses. 
Our goal is to find how the fraction of active synapses at resting conditions 
scales with brain size. In order to do this, we have first to
determine how $E$ scales with the brain volume. From Hofman (1983), we find
that a basal metabolic rate (an oxygen consumption by the whole cortex at rest)
$E\sim V_{g} V_{body}^{-0.15}$, where $V_{body}$ is the body volume. Using a 
scaling relation between the brain and body volumes $V_{g}\sim V_{body}^{0.75}$
(Hofman, 1983; Allman, 1999), we obtain $E\sim V_{g}^{0.8}$. On the other
hand, the total number of synapses in gray matter, i.e. the factor $N M$ 
in eq. (4.3) scales in first power with $V_{g}$ 
(Sch{\"u}z and Demianenko, 1995;
Braitenberg and Sch{\"u}z, 1998). From this, it follows that the fraction of
active excitatory synapses $f$ at resting conditions scales with brain size as 
$f\sim V_{g}^{-0.2}$, i.e there is a slow decay in value of $f$ as brain size 
increases. This result can suggest that metabolic energy is indeed the 
resource that brains try not to overuse.



\section{Appendix}


The degree of separation in a network composed of some units is defined 
as the minimal number of steps which are necessary to connect two arbitrary
units. 
The derivation of this quantity for cortical areas
is technically similar to the derivation of the average degree of
separation for neurons, which has been done elsewhere (Karbowski,
2001).  Here, we follow that approach.

A given two areas can be connected indirectly through many paths 
composed of chains of other areas in the cortex.
The probability $S_{n}$ that these two areas
are connected via at least one of the paths that uses $n$ steps
(i.e. the chain of $n$ intermediate areas) is given by 
(compare, Karbowski, 2001)

\set
\begin{equation}
S_{n}= 1 - (1-Q^{n})^{m_{n}},
\end{equation}\\
where $m_{n}= {K-2\choose n} n!\approx K^{n-1}$ is the number of all
possible paths that use exactly $n$ steps through which the two areas 
can be connected indirectly, and $Q$ is the average probability of 
a direct connection between cortical areas given by eq. (2.3) in the
main text. We can rewrite this formula in a more convenient form as 
we did in Sec. 2:

\set
\begin{equation}
S_{n}= 1 - e^{-p_{n}},
\end{equation}\\
where

\set
\begin{equation}
p_{n}= - K^{n-1}\ln(1-Q^{n}).
\end{equation}\\
For large $n$ we can expand the logarithm obtaining $p_{n}\approx
(KQ)^{n-1}Q$. Since $KQ$ is the average number of areas a cortical area
is connected to, it is substantially greater than $1$. As a consequence 
$p_{n} \gg 1$ as $n \gg 1$. We will use this fact later.

It is also instructive to determine how $p_{n}$ scales with the brain
size. From eq. (3.2) in the main text, we have to leading order in 
the Taylor expansion 
$Q\sim V_{g}^{-\delta}$ ($\delta= 0.28$),
and using another scaling of the number of areas $K$ with $V_{g}$, 
we obtain $p_{n}\approx K^{n-1}Q^{n}\sim V_{g}^{\alpha(n-1)-\delta n}= 
V_{g}^{2\beta n - \alpha(n+1)} $. Thus $p_{n}$ scales against $V_{g}$
with the exponent $2\beta n - \alpha (n+1)$. This implies that $p_{n}$
grows quickly with $V_{g}$ for $n > \alpha/(2\beta-\alpha)$, and decays
to zero for $n < \alpha/(2\beta-\alpha)$. In the former case,
from eq. (5.2) we get
$S_{n}\mapsto 1$ as $V_{g}\mapsto \infty$, i.e. there is a connection 
after $n$ steps in the limit of very big brains. 
In the latter case, this leads to $S_{n}\mapsto 0$ as 
$V_{g}\mapsto \infty$, i.e. there is no connection
at all. One can think about the boundary value 
$\alpha/(2\beta-\alpha)\approx 1.54$
as the average degree of separation between cortical areas in the limit 
of very big brains. As a comparison, the average degree of separation
between neurons is either 5, for smooth brains, or 10-11, for convoluted ones,
in this limit (Karbowski, 2001; Karbowski, 2002). 

Because we want to find the minimal number of steps which are needed
to connect indirectly two given areas,
we have to define a second probability $P_{n}$
that areas are connected in at least $n$ steps, i.e. $n$ is the minimal
number of necessary steps. These probabilities are related to the 
probabilities $\{ S_{n}\}$ in the following way

\set
\begin{eqnarray}
P_{n}= (1-S_{1})...(1-S_{n-1})S_{n} 
\end{eqnarray}\\
for $n \ge 2$, and $P_{1}= S_{1}$ for $n=1$. 
The average minimal number of steps, or equivalently, the average degree
of separation between cortical areas, defined as $<n>= \sum_{n=1}^{K-1} 
n P_{n}$, is given by

\set
\begin{eqnarray}
<n>= 2-Q + \sum_{j=2}^{K-2} \exp(-\sum_{i=1}^{j} p_{i})
- (K-1)\exp(- \sum_{i=1}^{K-1} p_{i}).
\end{eqnarray}\\
This exact but complicated equation can be approximated if we use the 
fact that $p_{n}$ rapidly becomes large as $n$ increases for a given $V_{g}$. 
This allows us to neglect all the exponents except the leading one, 
$\exp[-(p_{1}+p_{2})]$. Thus, approximately

\set
\begin{equation}
<n>\approx  2-Q + (1-Q)e^{-KQ^{2}},
\end{equation}\\
and it is apparent that the average degree of separation between areas
is close to $2$ with only minor species specific corrections and finite
size effects. For example, for rat with $K= 26$, $Q= 0.25$ we find 
$<n>= 1.89$, for cat with $K= 65$, $Q= 0.27$ we have $<n>= 1.73$, and for 
macaque monkey with $K= 73$, $Q=0.15$ we obtain $<n>= 2.01$.


\noindent{\bf Acknowledgments}

The author thanks Dr. Almut Sch{\"u}z for data on projection experiments 
in mouse and monkey, Bard Ermentrout, and an anonymous Reviewer for 
editorial comments on the manuscript.
The work was supported by NSF grant DMS 9972913, and by the Sloan-Swartz
fellowship at Caltech.

\vspace{1.5cm}

\noindent {\bf References} 

\noindent Allman, J.M. (1999). {\it Evolving Brains. \/} New York, Freeman.

\noindent Amir, Y., Harel, M., and Malach, R. (1993). Cortical hierarchy
reflected in the organization of intrinsic connections in macaque monkey
visual cortex. {\it J. Comp. Neurol., 334, 19-46. \/}

\noindent Attwell, D., and Laughlin, S.B. (2001). An energy budget for
signaling in the gray matter of the brain. {\it J. Cerebral Blood Flow
and Metabolism, 21, 1133-1145. \/}

\noindent Barton, R.A., and Harvey, P.H. (2000). Mosaic evolution of brain
structure in mammals. {\it Nature 405, 1055-1058. \/}

\noindent Braitenberg, V. (2001). Brain size and number of neurons: An
exercise in synthetic neuroanatomy.
{\it J. Comput. Neurosci., 10, 71-77. \/}

\noindent Braitenberg, V. (1978a). Cell assemblies in the cerebral cortex.
In {\it Theoretical approaches to complex systems. \/} Eds. R. Heim and
G. Palm. Berlin, Springer-Verlag.

\noindent Braitenberg, V. (1978b). Cortical architectonics: general and
areal. In {\it Architectonics of the cerebral cortex. \/} Eds. M.A.B. Brazier
and H. Petsche. New York, Raven, pp. 443-465.

\noindent Braitenberg, V., and Sch{\"u}z, A. (1998). {\it Cortex: Statistics
and Geometry of Neuronal Connectivity. \/} Berlin, Springer.

\noindent Burkhalter, A., and Bernardo, K.L. (1989). Organization of
corticocortical connections in human visual cortex.
{\it Proc. Natl. Acad. Sci. USA 86, 1071-1075. \/}

\noindent Cajal, S.R. (1995). {\it Histology of the Nervous System
of Man and Vertebrates. \/} New York, Oxford Univ. Press, vol.1.

\noindent Caviness, V.S. (1975). Architectonic map of neocortex of the normal
mouse. {\it J. Comp. Neurol., 164, 247-264. \/} 

\noindent Changizi, M.A. (2001). Principles underlying mammalian neocortical
scaling. {\it Biol. Cybern., 84, 207-215. \/}

\noindent Cherniak, C. (1995). Neural component placement.
{\it Trends Neurosci., 18, 522-527. \/}

\noindent Chklovskii, D. and Stevens, C.F. (1999). Wiring optimization
in the brain. In {\it Advances in
Neural Information Processing Systems. \/} Ed. S.A. Solla, 
Cambridge, MIT Press, vol. 12, pp. 103-107.

\noindent Douglas, R.J., and Martin, K.A.C. (1991). Opening the grey box.
{\it Trends Neurosci., 14, 286-293. \/}

\noindent Felleman, D.J., and Van Essen, D.C. (1991). Distributed hierarchical
processing in the primate cerebral cortex.
{\it Cereb. Cortex 1, 1-47. \/}

\noindent Frahm, H.D., Stephan, H., and Stephan, M. (1982). Comparison
of brain structure volumes in insectivora and primates. I. Neocortex.
{\it J. Hirnforsch., 23, 375-389. \/}

\noindent Greilich, H. (1984). Quantitative Analyse der cortico-corticalen
Fernverbindungen bei der Maus. {\it Thesis, Univ. of Tuebingen. \/}

\noindent Hofman, M.A. (1983). Energy metabolism, brain size and longevity
in mammals. {\it Quarterly Review of Biology, 58, 495-512. \/}

\noindent Hofman, M.A. (1985). Size and shape of the cerebral cortex
in mammals. I. The cortical surface. {\it Brain Behav. Evol. 27, 28-40. \/}

\noindent Hofman, M.A. (1989). On the evolution and geometry of the brain
in mammals. {\it Prog. Neurobiol. 32, 137-158. \/}

\noindent Hursh, J.B. (1939). Conduction velocity and diameter of nerve fibers.
{\it Amer. J. Physiol., 127, 131-139. \/}

\noindent Jerison, H.J. (1973). {\it Evolution of the brain and 
intelligence. \/} New York, Academic Press.

\noindent Jerison, H.J. (1991). {\it Brain size and the evolution of mind. \/}
New York, Am. Mus. Natl. Hist.

\noindent Kaas, J.H. (1987). The organization of neocortex in mammals:
Implications for theories of brain function. {\it Ann. Rev. Psychol.,
38, 129-151. \/}

\noindent Kaas, J.H. (1995). The evolution of isocortex. {\it Brain Behav.
Evol., 46, 187-196. \/}

\noindent Kaas, J.H. (2000). Why is brain size so important: Design
problems and solutions as neocortex gets bigger or smaller. {\it Brain
Mind, 1, 7-23. \/}

\noindent Karbowski, J. (2001). Optimal wiring principle and plateaus
in the degree of separation for cortical neurons. {\it Phys. Rev. Lett.,
86, 3674-3677. \/} 

\noindent Karbowski, J. (2002). Optimal wiring in the cortex and neuronal
degree of separation. {\it Neurocomputing, 44-46, 875-879. \/}

\noindent Kolb, B. (1990). Organization of the neocortex of the rat.
In {\it The cerebral cortex of the rat. \/} 
Eds. B. Kolb and R.C. Tees. Cambridge, MIT Press, pp. 21-33.

\noindent Krubitzer, L. (1995). The organization of neocortex in mammals:
Are species differences really so different? {\it Trends Neurosci.,
18, 408-417. \/}

\noindent Krubitzer, L., and Huffman, K.J. (2000). Arealization of the 
neocortex in mammals: genetic and epigenetic contributions to the 
phenotype. {\it Brain Behav. Evol., 55, 322-335. \/}

\noindent Krubitzer, L., Kunzle, H., and Kaas, J. (1997). Organization of
sensory cortex in Madagascan insectivore, the tenrec (Echinops telfairi).
{\it J. Comp. Neurol., 379, 399-414. \/}

\noindent Laughlin, S.B., de Ruyter van Steveninck, R.R., and Anderson,
J.C. (1998). The metabolic cost of neural information.
{\it Nature Neurosci., 1, 36-41. \/}

\noindent Levy, W.B., and Baxter, R.A. (1996). Energy-efficient neural codes.
{\it Neural Comput., 8, 531-543. \/}

\noindent Luhmann, H.J., Martinez-Millan, L., and Singer, W. (1986). 
Development of horizontal intrinsic connections in cat striate cortex.
{\it Exp. Brain Res., 63, 443-448. \/}

\noindent Mitchison, G. (1992). Axonal trees and cortical architecture.
{\it Trends Neurosci., 15, 122-126. \/}

\noindent Murre, J.M.J., and Sturdy, D.P.F. (1995). The connectivity of
the brain: multi-level quantitative analysis. {\it Biol. Cybern., 73, 
529-545. \/}

\noindent Olivares, R., Montiel, J., and Aboitiz, F. (2001). Species
differences and similarities in the fine structure of the mammalian
corpus callosum. {\it Brain Behav. Evol., 57, 98-105. \/}

\noindent Pandya, D.N., and Yeterian, E.H. (1985). Architecture and 
connections of cortical association areas.
In {\it Cerebral Cortex. \/} 
Eds. A. Peters and E.G. Jones. New York, Plenum, vol. 4,  pp. 3-61.

\noindent Prothero, J.W., and Sundsten, J.W. (1984). Folding of the cerebral
cortex in mammals. {\it Brain Behav. Evol. 24, 152-167. \/}

\noindent Ringo, J.L. (1991). Neuronal interconnection as a function of 
brain size. {\it Brain Behav. Evol., 38, 1-6. \/} 

\noindent Ringo, J.L., Doty, R.W., Demeter, S., and Simard, P.Y. (1994). 
Time is of the essence: A conjecture that hemispheric specialization arises
from interhemispheric conduction delay.
{\it Cereb. Cortex 4, 331-343. \/} 

\noindent Rockel, A.J., Hiorns, R.W., and Powell, T.P.S. (1980).
The basic uniformity in structure of the neocortex. {\it  Brain 103,
221-244. \/} 

\noindent Rockland, K.S., Lund, J.S., and Humphrey, A.L. (1982). 
Anatomical binding of intrinsic connections in the striate cortex of
tree shrews (Tupaia glis). {\it J. Comp. Neurol., 209, 41-58. \/}

\noindent Rumberger, A., Tyler, C.J., and Lund, J.S. (2001). Intra- and 
inter-areal connections between the primary visual cortex V1 and the area
immediately surrounding V1 in the rat.
{\it Neuroscience 102, 35-52. \/}

\noindent Rushton, W.A.H. (1951). A theory of the effects of fiber size in
medullated nerve. {\it J. Physiol., 115, 101-122. \/}

\noindent Scannell, J.W., and Young, M.P. (1993). The connectional
organization of neural systems in the cat cerebral cortex.
{\it Current Biology 3, 191-200. \/}

\noindent Scannell, J.W., Young, M.P., and Blakemore, C. (1995). Analysis
of connectivity in the cat cerebral cortex.
{\it J. Neurosci., 15, 1463-1483. \/}

\noindent Sch{\"u}z, A. (2001), private communication.

\noindent Sch{\"u}z, A., and Demianenko, G. (1995). Constancy and variability 
in cortical structure. A study on synapses and dendritic spines in hedgehog
and monkey. {\it J. Brain Res., 36, 113-122. \/}

\noindent Sch{\"u}z, A., and Liewald, D. (2001). Patterns of cortico-cortical 
connections in the mouse. 
In {\it The neurosciences at the turn of the century. \/} 
Eds. N. Elsner and G.W. Kreutzberg. Thieme, vol. II, p. 624.

\noindent Sch{\"u}z, A., and Preissl, H. (1996). Basic connectivity of the 
cerebral cortex and some considerations on the corpus callosum. 
{\it Neurosci. Biobehav. Rev., 20, 567-570. \/}

\noindent Shulman, R.G., and Rothman, D.L. (1998). Interpreting functional
imaging studies in terms of neurotransmitter cycling. {\it Proc. Natl. Acad.
Sci. USA, 95, 11993-11998. \/}

\noindent Sibson, N.R., Dhankar, A., Mason, G.F., Rothman, D.L., Behar, K.L.,
and Shulman, R.G. (1998). Stoichiometric coupling of brain glucose metabolism
and glutamatergic neuronal activity. {\it Proc. Natl. Acad. Sci. USA, 95, 
316-321. \/}

\noindent Sporns, O., Tononi, G., and Edelman, G.M. (2000). Theoretical
Neuroanatomy: Relating anatomical and functional connectivity in graphs
and cortical connection matrices.
{\it Cereb. Cortex, 10, 127-141. \/}

\noindent Stephan, H., Frahm, H., and Baron, G. (1981). New and revised
data on volumes of brain structures in insectivores and primates.
{\it Folia Primatol. (Basel), 35, 1-29. \/}

\noindent Stevens, C.F. (1989). How cortical interconnectedness varies
with network size. {\it Neural Comput. 1, 473-479. \/}

\noindent Wong-Riley, M.T.T. (1989). Cytochrome oxidase: an endogenous
metabolic marker for neuronal activity. {\it Trends Neurosci., 12, 94-101. \/}

\noindent Young, M.P. (1992). Objective analysis of the topological
organization of the primate cortical visual system.
{\it Nature 358, 152-155. \/}

\noindent Young, M.P. (1993). The organization of neural systems in
the primate cerebral cortex.
{\it Proc. Roy. Soc., B 252, 13-18. \/}

\noindent Young, M.P., Scannell, J.W., and Burns, G. (1995). 
{\it The analysis of cortical connectivity. \/} Austin, TX, Landes.

\noindent Zeki, S, and Shipp, S. (1988). The functional logic of cortical
connections. {\it Nature 335, 311-317. \/}

\noindent Zhang, K., and Sejnowski, T.J. (2000). A universal scaling law 
between gray matter and white matter of cerebral cortex. 
{\it Proc. Natl. Acad. Sci. USA 97, 5621-5626. \/}

\newpage

{\bf \large Figure Captions}

Fig. 1\\
Dependence of the average size of cortical patches in V1 on the logarithm of
the gray matter volume $V_{g}$ for several mammals. Data for the patches 
widths are:
for rat $\xi= 370$ $\mu m$ (Rumberger et al 2001), for
tree shrew $\xi= 230$ $\mu m$ (Rockland et al 1982), for cat 
$\xi= 300$ $\mu m$ (Luhmann et al 1986), for macaque $\xi= 230$ $\mu m$
(Amir et al 1993), for human $\xi= 400$ $\mu m$ (Burkhalter and Bernardo
1989). The gray matter volumes were taken form Stephan et al, 1981.

\vspace{0.3cm}

Fig. 2\\
Log-log dependence of the average axon length $L_{0}$ in white matter
on the gray matter volume $V_{g}$ in one hemisphere for few mammals: 
mouse, rat, cat, and macaque. We find the allometric law
$L_{0}= 1.10\ V_{g}^{\beta}$. The scaling exponent 
$\beta= 0.26\pm 0.02$. Data for the axon length come from
computation performed in Sec. 2. For macaque, it was taken $L_{0}= 14.5$
mm as an average value between 12.0 and 17.0 mm. The gray matter volumes 
were taken form Stephan et al (1981) and rescaled by 0.5.

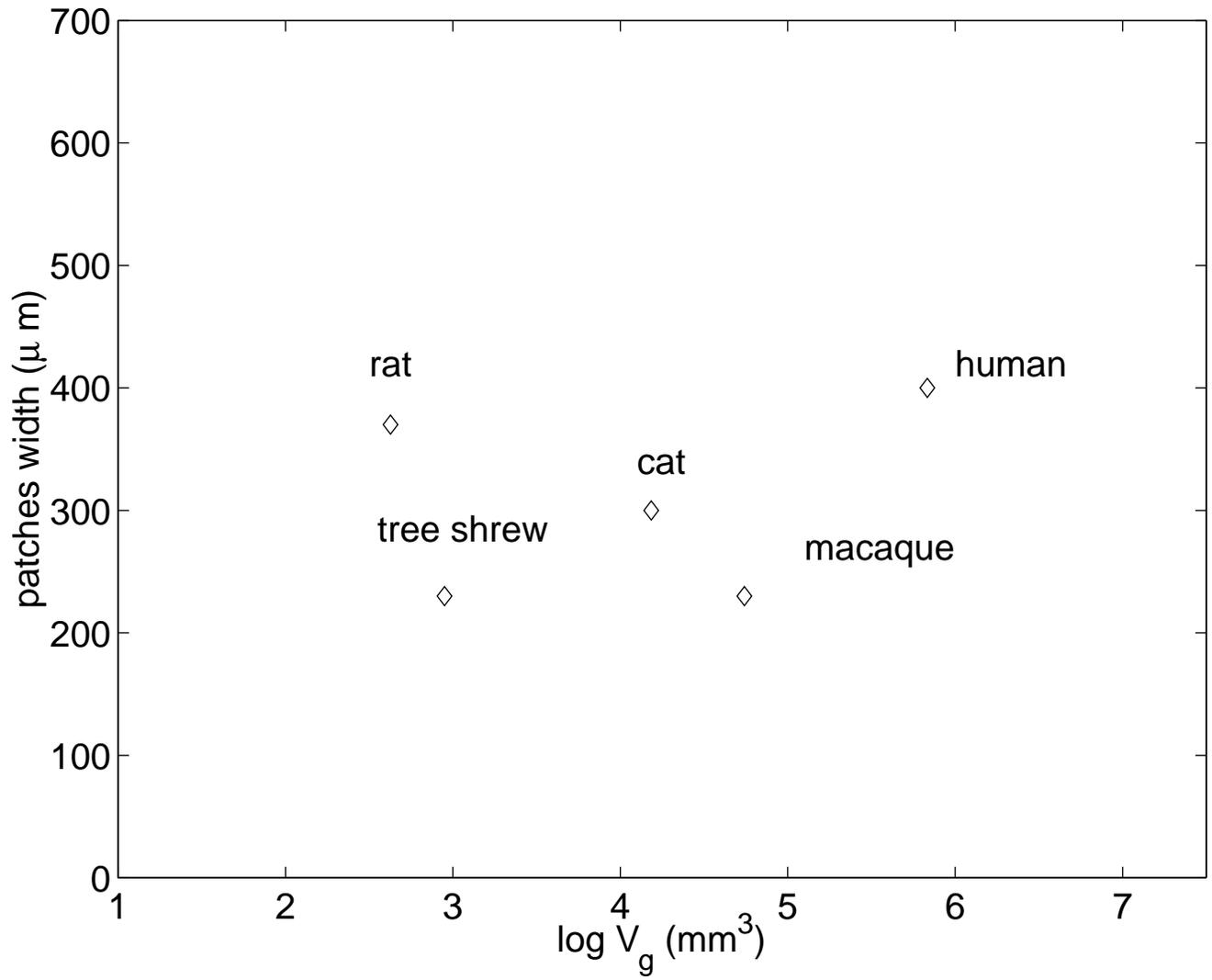
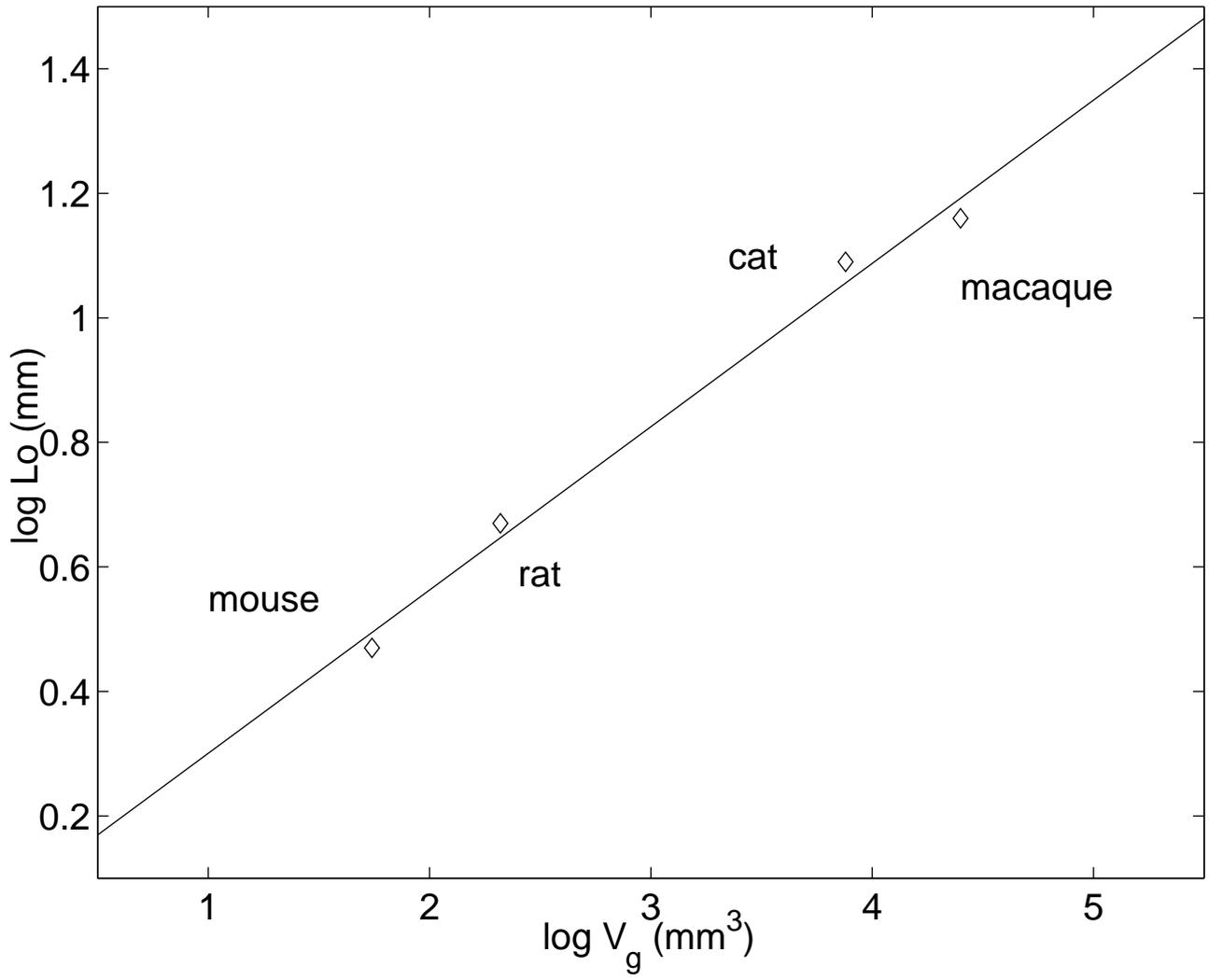
\end{document}